\documentclass[aps,twocolumn,amsmath,amssymb,floatfix,showpacs]{revtex4}
\usepackage{graphicx,color, bm} 
\usepackage{dcolumn} 

\begin{document}
\title{
Conformal invariance in three-dimensional rotating turbulence}
\author{S. Thalabard$^{1,2}$, D. Rosenberg$^1$,  A. Pouquet$^1$ and P.D. Mininni$^{1,3}$}
\affiliation{
 $^1$Computational and Information Systems Laboratory, NCAR, 
         P.O. Box 3000, Boulder CO 80307, USA. 
$^2$ CEA Saclay, l'Orme les Merisiers, France.\\
$^3$Departamento de F\'\i sica, Facultad de Ciencias Exactas y
         Naturales, UBA \& IFIBA, CONICET, Ciudad 
         Universitaria, 1428 Buenos Aires, Argentina. 
             }
\date{\today}
\begin{abstract}
We examine  three--dimensional turbulent flows in the presence  of solid-body rotation and helical forcing in the framework of stochastic Schramm-L\"owner evolution curves (SLE). The data stems from a run on a grid of $1536^3$ points, with Reynolds and Rossby numbers of respectively 5100 and $0.06$. We average the parallel component of the vorticity in the direction parallel to that of rotation, and examine the resulting $\left<\omega_\textrm{z}\right>_\textrm{z}$ field for scaling properties of its zero-value contours. We find for the first time for three-dimensional fluid turbulence evidence of nodal curves being conformal invariant, belonging to a SLE class with associated Brownian diffusivity $\kappa=3.6\pm 0.1$. SLE behavior is related to the self-similarity of the direct cascade of energy to small scales in this flow, and to the partial bi-dimensionalization of the flow because of rotation.  We recover the value of $\kappa$ with a heuristic argument and show that this value is consistent with several non-trivial SLE  predictions.
\end{abstract}
\maketitle

Self-similarity in physics is a common phenomenon, with identical properties of a system when considered at different scales. Rugged coast lines, fractals, traffic in computer networks, growth processes, geometrical properties of interfaces, phase transitions in critical phenomena such as in the Ising model for spontaneous magnetization, classical and quantum field theory,  often display power-law scaling of some variable and such scaling exponents have been the object of intense investigations resulting in the finding of broad classes of universality. 

A property stronger than scale invariance is conformal invariance, under transformations that preserve angles with rescaling that depends on position; it is difficult to test, since it implies the need to investigate the scaling of multi-point high-order correlation functions. However, recent developments by Schramm in particular (see e.g. \cite{revs} and references therein) allow in some cases for a statistical characterisation of conformal invariance. Such scaling laws can be related to Brownian motion (which is scale invariant, and conformal in two dimensions) in what is now named Schramm-L\"owner evolution (SLE), with as sole parameter the diffusivity $\kappa$ associated with this Brownian motion. In this approach, the driving of the L\"owner equation (Eq.~\ref{LOW} below) is stochastic, with a conformal map allowing to go from static (fixed time) two-dimensional (2D) paths in the complex plane $\mathbb{C}$ to ``dynamic'' one-dimensional (1D) motions. In other words, it allows one to describe paths in $\mathbb{C}$ by a succession (convolution) of conformal maps obeying a differential equation. Schramm's theorem (see, e.g., \cite{revs}) states that if and only if the driving is Brownian is the measure of the 2D paths conformally invariant.

Two-dimensional turbulence differs in many ways from the  three-dimensional (3D) case because of the presence of an extra invariant  in the absence of viscosity, the enstrophy $S=\left<|\nabla \times {\bf u}|^2/2\right>$, leading to an inverse cascade of energy $E=\left<|{\bf u}|^2/2\right>$ to large scales \cite{rhk_montgo}, with ${\bf u}$ the velocity field. It was shown in \cite{Bernard06} that  this inverse cascade, which is known to lack intermittency and is self-similar, can be viewed in the framework of conformal invariance when examining zero-vorticity lines; it belongs to the universality class with $\kappa=8/3$ (the enstrophy cascade to small scales, however, is not SLE  \cite{Bernard06}). These results stem from direct numerical simulations (DNS) on grids of up to $16,384^2$ points, with forcing at intermediate wavenumber, $k_F/k_{min}\approx 100$, with $k_{min}=2\pi/L_0$, $L_0$ being the size of the vessel.

In the case of 3D Navier-Stokes (NS) incompressible flows at high Reynolds numbers, the cascade of energy to small scales is not self-similar, because of the presence of strong vorticity gradients. Only one time scale is present, the eddy turn-over time $\tau_\textrm{NL}\sim \ell/u_{\ell}$, with $u_{\ell}$ the velocity at scale $\ell$, and dimensional analysis gives an energy spectrum $E(k)\propto k^{-5/3}$ that is quite close to observed spectra in the atmosphere or in laboratory experiments. However, when introducing solid body rotation $\Omega$ with inertial time $\tau_\Omega \sim 1/\Omega$, $E(k)$ steepens and its  spectral index can be recovered by taking into account the weakening of nonlinear interactions due to the inertial waves \cite{Cambon}. In this case, self-similarity and Gaussianity in the 3D direct energy cascade was found recently both in the laboratory \cite{pinton, van} and in DNS \cite{rot512, 1536a}, more clearly so in the presence of helicity, i.e., velocity-vorticity correlations \cite{1536b}. 
 
Since rotating flows tend to become quasi-2D (but not strictly 2D, as our results will confirm) when strong rotation is imposed, the question thus arises as to whether SLE  can be identified in such flows. To this end, we examine the large data set produced in a run of rotating helical turbulence on a grid of $1536^3$ points, with $L_0=2\pi$ and forcing at $k_F$=7; an inverse cascade of energy to large scales (with constant negative flux) 
is observed, but with too little extent in wavenumber to allow for a SLE  analysis similar to that performed in \cite{Bernard06} for the 2D NS inverse cascade. We concentrate instead on the direct energy cascade to small scales (with constant and positive flux, see Fig. 9 in \cite{1536a}), and analyzed for its classical statistical properties and structures in \cite{1536b}. A pseudo-spectral code with periodic boundary conditions was used, with at the onset of the inverse cascade a Reynolds number $Re=U_0 2\pi/[\nu k_F] \approx 5100$ (with $\nu$ the viscosity), and the Rossby number $Ro= U_0k_F/[2\pi \Omega] \approx 0.06$; $U_0\approx 1$ is the {\it r.m.s.}~velocity. We integrated the 3D NS equations in the rotating frame for an incompressible flow ($\nabla \cdot {\bf u} =0$); with $\mbox{\boldmath $\omega$} = \nabla \times {\bf u}$ the vorticity, they read:
\begin{equation}
\frac{\partial {\bf u}}{\partial t} + \mbox{\boldmath $\omega$} \times
    {\bf u} + 2 \mbox{\boldmath $\Omega$} \times {\bf u}  =
    - \nabla {\cal P} + \nu \nabla^2 {\bf u} + {\bf F} \ ;
\label{eq:momentum}
\end{equation}
${\cal P}$ is the total pressure  modified by the centrifugal term, and {\bf F} is a helical Arn'old-Beltrami-Childress forcing \cite{rot512, 1536a}. The rotation is imposed in the vertical (z) direction, with ${\bf \Omega}=9 \hat{\textrm{z}}$. The code is fully parallelized, uses the $2/3$ de-aliasing rule, and the temporal scheme is a second-order Runge-Kutta. Note that in 3D, besides energy, total helicity $H=\left<{\bf u} \cdot \mbox{\boldmath $\omega$} \right>$ is also an ideal invariant \cite{Moffatt69}.

\begin{figure}
\hskip-0.15truein \vskip-0.17truein \includegraphics[width=0.997\linewidth]{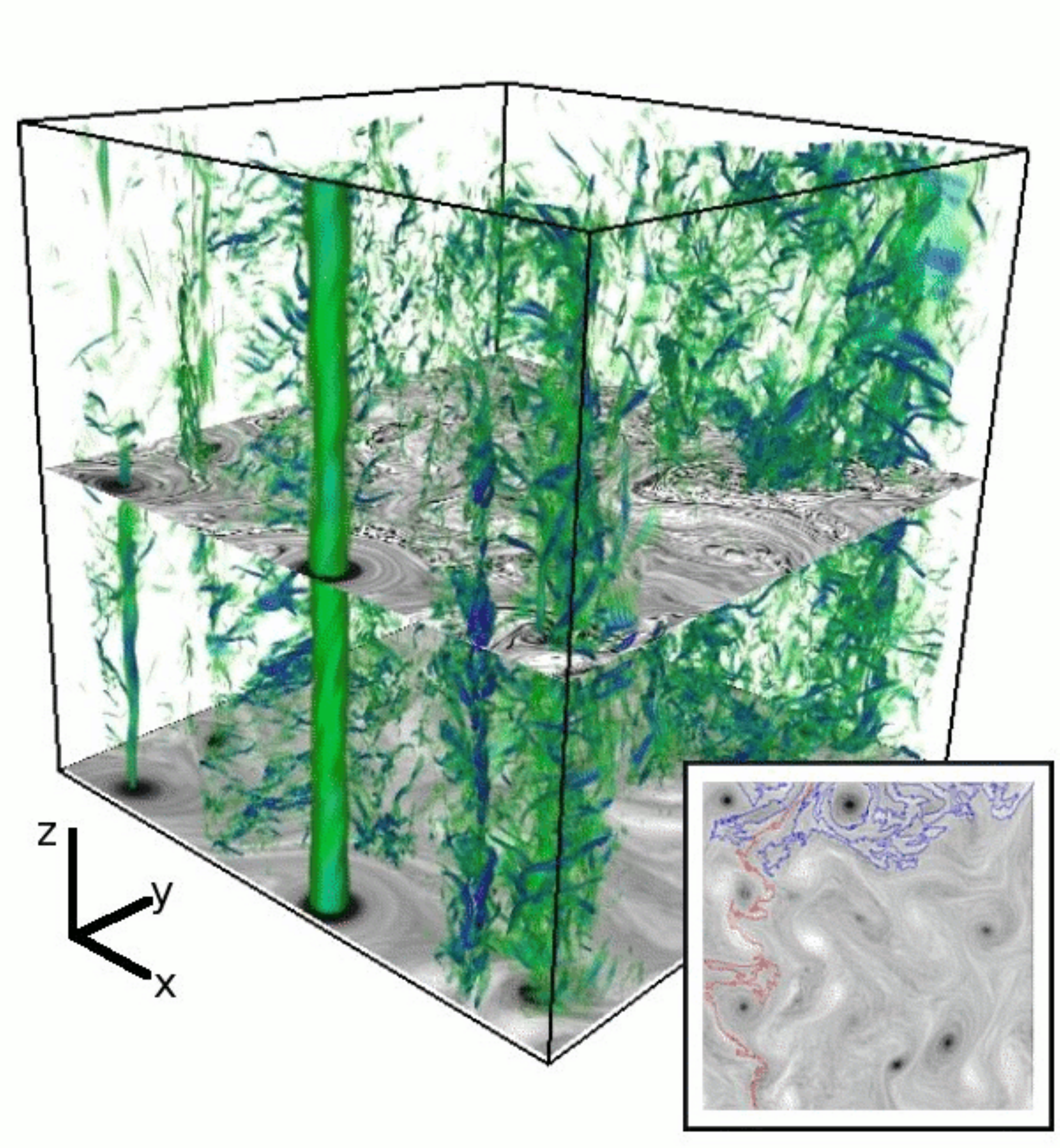}
\caption{({\it Color online}) Perspective volume rendering of vorticity intensity in a snapshot of the flow. The slice in the middle of the box shows a 2D cut of $\omega_\textrm{z}$, while the slice at the bottom shows the same field averaged vertically. The inset is $\left<\omega_\textrm{z}\right>_\textrm{z}$, with super-imposed nodal paths, the traversing ones (top to bottom, red online) being discarded from the analysis.}
\label{pacman_1536}
\end{figure}

{\it The procedure:}
Considering the symmetries of Eq.~(\ref{eq:momentum}), we construct a 2D field by averaging in the vertical direction the parallel component of vorticity, which we denote hereafter $\left<\omega_\textrm{z}\right>_\textrm{z}$; we also compute a transverse average $\left<\omega_\textrm{z}\right>_\textrm{y}$ to compare with. Starting from an arbitrary line, say the $x$-axis, we explore iso-contours of zero field as trajectories in the 2D plane that keep positive field to their right. The direction along the trajectories is parametrized by a ``driving time'' $\tau$. The path is stopped whenever it returns to the initial axis. The end-point is then sent to infinity through a holomorphic (M\"obius) transformation as in \cite{Bernard06}, with a cut-off $\Delta$ chosen to be such that the tip of the curve is within a small arbitrary distance of the chosen axis; results are insensitive to the choice of $\Delta$ in a range of 1 to 10 pixels, and agree as well with a procedure in which the M\"obius conformal mapping is not applied. Note that, because of periodicity, the procedure is not affected by the boundaries, and that all trajectories are renormalized to $\tau_{max}=1$.

We have shown numerically for this flow the existence of scale invariance for the direct energy cascade and the Gaussianity of the velocity in \cite{1536b} (see Figs. 7 and 8), also examining anisotropy at different times using a $SO(2)\times \mathbb{R}$ decomposition (see Figs. 2 and 3 in \cite{1536b} for the actual scaling ranges). We now probe the conformal invariance of these 2D curves viewed as paths in the upper complex plane; the paths are encoded in a ``driving function'' $\xi(\tau)$ obtained through the chordal L\"owner equation below, with $g_{\tau}(\zeta)$ ($\zeta \in \mathbb{C}$) a conformal map (see, e.g., \cite{Kennedy}):
\begin{equation}
\partial_{\tau} g_{\tau}(\zeta) =\frac{2}{g_{\tau}(\zeta)-\xi(\tau)} \ ;
\label{LOW} \end{equation}
$\xi(\tau)$ is the unknown 1D real continuous stochastic driving function for the path. In order to estimate $\xi(\tau)$ numerically, we use the zipper algorithm (ZA) with vertical slits \cite{MARSHALL}.  Then 
$g_{a,\delta\tau}(\zeta) = a + \sqrt{{[(\zeta-a)^2 +  4 \delta \tau]}}$
conformally maps the upper plane minus the vertical slit in $\mathbb{C}$, $[(a,0) ; (a, 2\sqrt\tau)]$, into the upper plane: ZA  gradually  zips the whole path onto the $x$-axis using the composition of functions $g_{a,\delta\tau}(\zeta)$ for different $\delta \tau$. We thus transform the erratic nodal line in the plane (inset in Fig. \ref{pacman_1536}, described below) into an unknown motion along the real axis, $\xi(\tau)$.

To test for conformal invariance, we therefore must ask: Is $\xi(\tau)$ a Gaussian process? Does it correspond to a Brownian motion? And if so, what is its diffusivity? To answer the first question, one can use the classical Kolmogorov-Smirnov (KS) test, and check (i) whether its $p_{KS}$ value is above a given threshold for a wide range of driving times $\tau$, and (ii) whether the steps in this motion are independent. When both tests are favorable, we then consider the scaling of the variance of $\xi(\tau)$. If the scaling is reasonably linear with $\tau$, we will conclude that the set of driving functions likely stems from a Brownian process, and hence that the vorticity isolines obtained as indicated above are likely to be conformally invariant. The linear scaling also gives us the diffusivity $\kappa$ which describes entirely the statistics of the SLE process.

\begin{figure}
\includegraphics[trim=20 20 0 0,width=0.93\linewidth]{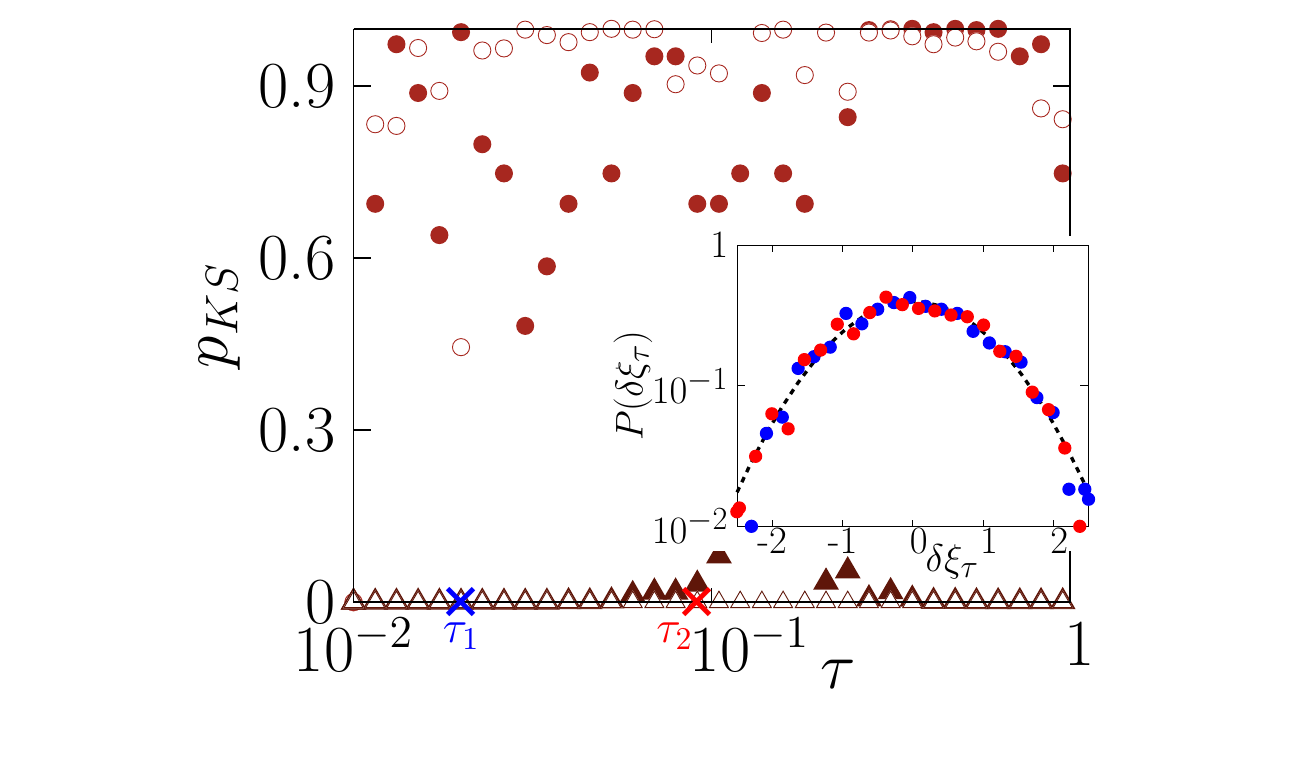} \includegraphics[trim=4 15 0 0,width=0.80\linewidth]{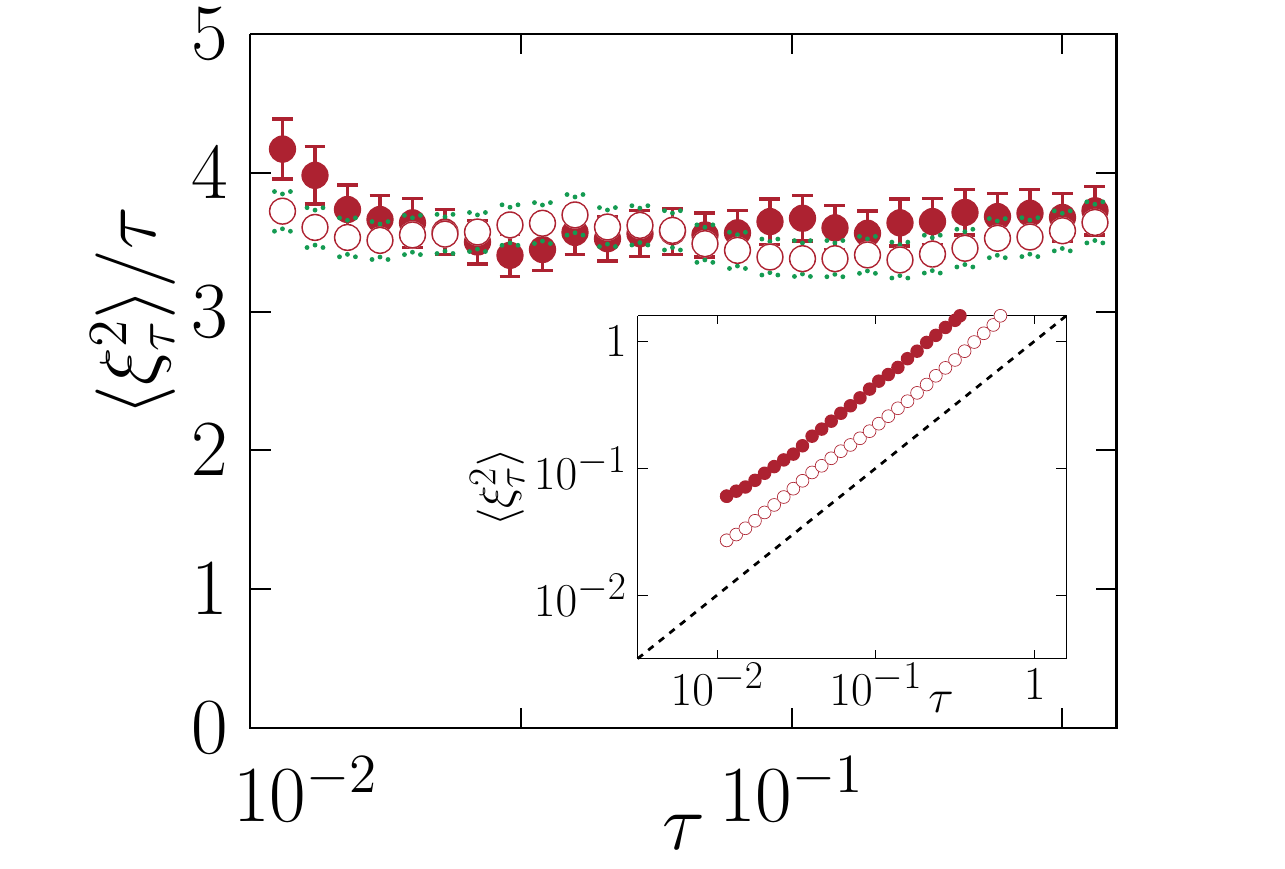}
\caption{({\it Color online}) SLE  analysis for $\left< \omega_\textrm{z} \right>_\textrm{z}$ (circles) and $\left< \omega_\textrm{z} \right>_\textrm{y}$ (triangles). Filled symbols comprise the full data, and open symbols correspond to paths in the dataset filtered at $\ell< 2\pi/10$. Driving time $\tau$ is in log coordinates. 
{\it Top:} KS test; inset: PDFs of driving function increments for $\left< \omega_\textrm{z} \right>_\textrm{z}$ for $\tau_1$ and $\tau_2$ as marked on the $\tau$--axis; the dashed line indicates a Gaussian. 
{\it Bottom:} Diffusivity $\kappa$ for $\left<\omega_\textrm{z}\right>_\textrm{z}$ 
with error bars; inset: scaling of variance for $\left< \omega_\textrm{z} \right>_\textrm{z}$, with dash line indicating linear variation with $\tau$.}
\label{kf30_1536} \end{figure}

{\it Results:} 
We now apply the procedure to the $1536^3$ DNS data. After performing the average (either in z or in y), fifteen temporal snapshots are analyzed, separated by approximately one eddy turn-over time. The resulting dataset has in excess of $3.5\times 10^7$ points for each averaging direction. In Fig.~\ref{pacman_1536} is given a snapshot of $|\mbox{\boldmath $\omega$}({\bf x})|$ in 3D, a 2D slice of $\omega_\textrm{z}({\bf x})$ (in the middle of the box), and the same field component when vertically averaged, $\left<\omega_\textrm{z}\right>_\textrm{z}$ (bottom slice). The flow displays features of both 2D and 3D behavior \cite{rot512, 1536b}; prominent are tangles of vortex filaments loosely organized parallel to $\Omega \hat z$, and Beltrami core vortices, smooth long-lived helical columns.  The inset shows $\left<\omega_\textrm{z}\right>_\textrm{z}$ face-on with a few examples of superimposed paths that are analyzed below, noting that we discard the traversing paths, only keeping returning paths (loops at the top, blue online), as done in \cite{Bernard06,SQG}.

Figure \ref{kf30_1536} summarizes the analysis, for $\omega_\textrm{z}$ averaged either parallel (full circle) or transverse (y, full triangle) to $\Omega \hat{\textrm{z}}$. Since a barely-resolved inverse cascade of energy develops in the DNS between the forcing scale and the box size, we also performed the analysis in a dataset in which $\omega_\textrm{z}$ was filtered so as to only preserve scales smaller than the driving scale: for $\ell < 2\pi/10$, the results are now displayed with open symbols. Figure \ref{kf30_1536} (top) gives the $p_{KS}$ values of the KS test with abscissa $\tau$ in log scale. The value $p_{KS} \approx 10^{-5}$ for $\left< \omega_\textrm{z} \right>_\textrm{y}$ (triangles) shows that the transverse y-averaged field is not Gaussian, and we shall not analyze further such y-averaged data. On the other hand, $p_{KS} \ge 0.6$ for most values of $\tau$ for $\left<\omega_\textrm{z}\right>_\textrm{z}$ (circles). These opposite results imply that our test can eliminate non-Gaussian behavior, and that, due to the anisotropy of the flow, only parallel z-averaging may lead to conformal invariance. To confirm the Gaussianity of the process with {\sl parallel} averaging, we show in the inset the probability distribution functions (PDFs) for two driving times $\tau_1$ and $\tau_2$; the dotted line is a Gaussian with zero mean and unit variance \cite{note2}. Note that Gaussianity also implies independence of increments (we show the evolution of the PDFs for different $\tau$ in Fig. \ref{kf30_1536}).

\begin{figure} 
\includegraphics[trim=30 20 0 0,width=1.05\linewidth]{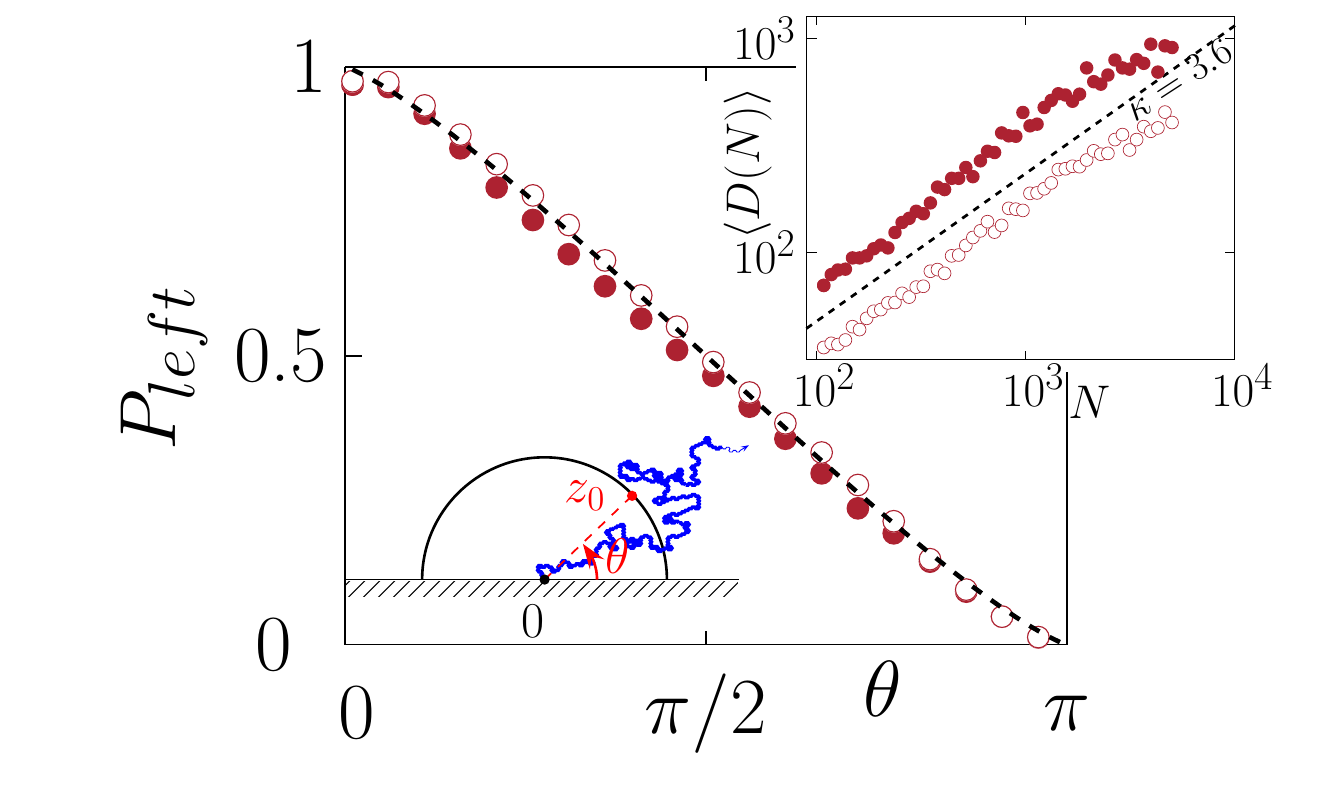} 
\caption{({\it Color online}) Winding angle probability $P_\textrm{\it left}$ as a function of the angle $\theta$, as illustrated by the sketch at the bottom.
{\it Inset:} Mean gyration radius $D$ of nodal lines as a function of the number of pixels $N$. Dashed lines: theoretical predictions for $\kappa_S = 3.6$; open and filled circles as in Fig.~\ref{kf30_1536}.
} \label{kf30_1536-2}
\end{figure}

Figure \ref{kf30_1536} (bottom) gives the variance of the motion normalized by $\tau$ with error bars for the datasets that are not discarded by the KS test. The resulting diffusivity for the associated Brownian motion is  $\left< \xi^2_{\tau}\right>/\tau=\kappa= 3.6 \pm 0.1$ for the full data set (full circles), and $\kappa=3.5\pm 0.2$ for the data in which the scales comparable to or larger than the forcing have been filtered out (open circles).
The inset shows the actual scaling of variance with driving time, in log-log coordinates, with the dashed line for linear variation. Note that, within error bars, the results are insensitive to whether or not we filter the numerical data (keeping only Fourier modes $k>10$, thereby making sure we restrict the data to the direct cascade of energy).

Finally, we confirm the scaling we found for $\kappa$ by examining some of the predictions on statistical properties of nodal lines of $\left< \omega_\textrm{z} \right>_\textrm{z}$ that can be made using the SLE framework (see \cite{revs}). A classical one concerns the fractal dimension of the nodal lines, but less trivial features predicted by SLE include for instance the so-called ``winding angle'' or the gyration radius. The winding angle prediction states that the probability $P_\textrm{\it left}$ of a SLE line to leave a point $z_0 = \rho e^{i\theta}$ in $\mathbb{C}$ to its left depends only on $\kappa$ and $\theta$ following a known expression \cite{revs}. Figure \ref{kf30_1536-2} shows the results obtained from our datasets as a function of $\theta$, as well as the mean gyration radius of the nodal lines as a function of their length in pixels in the top-right inset. In both cases, the SLE  predictions for $\kappa=3.6$, given with the dashed lines, appears convincing.

We thus conclude that our analysis identifies conformal invariance for nodal lines of the vertical component of the vorticity field when averaging parallel to the direction of rotation, and fails to identify such invariance in its transverse average. For the parallel-averaged vertical vorticity, the associated diffusivity is $\kappa\approx 3.6\pm 0.1$. Moreover, SLE  predictions for this value of $\kappa$ agree well with our results. It is also important to remark that our analysis could fail to reject the hypothesis of Gaussianity if data were insufficiently resolved; this is not surprising since it is hard to distinguish SLE  behavior from something close to SLE \cite{MARSHALL}. In spite of these limitations, the data analyzed here up to the spatial resolution considered is found to be consistent with SLE behavior.

{\it Discussion:} 
Rotating helical turbulence may be perhaps the first documented case presenting SLE  scaling for three-dimensional flows undergoing a direct cascade of energy and of helicity to small scales, when properly averaged in the direction of rotation. Conformal invariance is a strong local property and allows determination of a series of scaling laws, as exemplified in \cite{Bernard06,SQG} for 2D NS and other related 2D cases such as surface quasi-geostrophic (SQG) flows, and as found here as well. SLE  obtains convincingly for the vertical component of the vorticity averaged along the direction of rotation, with $\kappa\approx 3.6$, close (but not identical) to the value identified in \cite{SQG} for SQG flows for an inverse cascade. Note that anisotropy of this 3D rotating flow must play an essential role, since the direct cascade of enstrophy in strictly 2D NS is not SLE as shown in \cite{Bernard06}.

The fractal dimension D$_F\le 2$ of SLE  curves can be related to $\kappa$ \cite{Bernard06}, as well as to the cancellation exponent $\kappa_C$ which measures how fast a field changes sign \cite{luca}. With $d$ the dimension of space, we have $D_F=1+\kappa/8 = d-2\kappa_C$. It is straightforward to relate the diffusivity of the SLE process $\kappa$ and the exponent $e$ of the energy spectrum, $E(k)\sim k^{-e}$, under the assumption of self-similarity, $\zeta_p=a_Sp$; $\zeta_p$ are the exponents of the $p$th-order longitudinal structure functions of the velocity field, $\left<\delta u_L(r)^p\right>\sim r^{\zeta_p}$, where $\delta u_L$ is the variation of the velocity projected along the direction of the spatial increment ${\bf r}$. We use that  $\kappa_C=\zeta_1=a_S$, $\zeta_2=e-1$ for $1<e\le 3$, and that dimensional analysis for a given dynamics gives $a_S$. Then,
\begin{equation}
\frac{\kappa}{8} = d-e  = d-1-2a_S \ .
 \label{phenomeno}   \end{equation}
Hence, the value of $\kappa$ is quite sensitive to $e$ or $a_S$ \cite{note3}. For 2D NS, $a_S$=1/3 and $\kappa$=8/3, as found  in \cite{Bernard06} (with dual value $\kappa^{*}=6$). For rotating helical turbulence, $a_S$=3/4, using a phenomenological model based on three assumptions \cite{rot512, 1536b}: wave-modulated energy spectrum; domination of the helicity cascade to small scales; and maximal helicity. The first hypothesis allows to write that the transfer of energy to small scale is slowed down in the proportion $\tau_\textrm{NL}/\tau_\Omega$; the second one stems from the fact that the energy undergoing an inverse cascade to large scale, little energy is left to feed the small scales, whereas helicity only possesses a small-scale cascade and thus is the determining factor in this direct cascade. These two concepts lead to $e+h=4$, with helicity spectrum $H(k)\sim k^{-h}$. The third assumption gives $h=e-1$ and thus $\zeta_p=3p/4$, a value reported experimentally as well \cite{van, note3}. From Eq.~(\ref{phenomeno}), we then obtain $\kappa=4$, close to the value we find given the statistics. 

The connection between SLE  and statistical properties of turbulence allows one to look at such flows with a new eye, and to build bridges between fluid dynamics and other research areas in mathematics, condensed matter, percolation, and quantum field theory. Other three-dimensional flows may be studied with the same tools when the flow is self-similar and symmetries allow for a reduction of dimensionality. As an example, we leave for future work an investigation of SLE  properties in the inverse cascade of rotating turbulence.

\noindent {\it Computer time was provided by NCAR, which is sponsored by NSF; partial support was given by grants NSF-CMG 1025188, and for PDM by UBACYT 20020090200692, PICT 2007-02211, and PIP 11220090100825.}


\begin{thebibliography}{10}
\bibitem{note2}
A similar test for $\langle \omega_\textrm{z} \rangle_\textrm{y}$
gives strong departures from a Gaussian for its driving function increments  (not shown). 
\bibitem{note3}
Different problems with the same inertial index $e$ can lead to the same diffusivity $\kappa$; conversely, a set of rotating fluids with different energy indices ($e+h=4, e\not= h$) can lead to different diffusivities, provided they are SLE.
\end{thebibliography}
\end{document}